\documentclass{aa}
\usepackage{txfonts}
\usepackage{psfig} 

\def\pks{PKS~0558-504\/}
\def\ros{{\sl ROSAT }}
\def\asca{{\sl ASCA }}
\def\ergs{erg\,s$^{-1}$}

\newcommand{\ltsima} {$\; \buildrel < \over \sim \;$}
\newcommand{\gtsima} {$\; \buildrel > \over \sim \;$}
\newcommand{\lta} {\lower.5ex\hbox{\ltsima}}
\newcommand{\gta} {\lower.5ex\hbox{\gtsima}}

\def\approxlt{\mathrel{\hbox{\rlap{\lower.55ex \hbox {$\sim$}}
        \kern-.3em \raise.4ex \hbox{$<$}}}}
\def\approxgt{\mathrel{\hbox{\rlap{\lower.55ex \hbox {$\sim$}}
        \kern-.3em \raise.4ex \hbox{$>$}}}}

\begin{document}

\title{X-ray variability of the Narrow Line Seyfert~1 
       Galaxy PKS~0558$-$504}  
\author{W. Brinkmann\inst{1}  \and P. Ar\'evalo\inst{2}
  \and M. Gliozzi\inst{3} \and E. Ferrero\inst{2}}
\offprints{W. Brinkmann; e-mail: wpb@mpe.mpg.de}
\institute{Centre for Interdisciplinary Plasma Science, 
Max-Planck-Institut f\"ur extraterrestrische Physik,
         Postfach 1603, D-85740 Garching, Germany
\and Max-Planck-Institut f\"ur extraterrestrische Physik,
         Postfach 1603, D-85740 Garching, Germany
\and George Mason University,
Department of Physics and Astronomy, MS 3F3, 4400 University Dr., Fairfax, 
VA 22030-4444
}
\date{Received: 21.07.03 ; accepted: 20.11.03 }

\abstract{
We present results from several XMM-Newton observations
of the radio loud Narrow-line Seyfert 1 galaxy (NLS1) PKS~0558-504.
We find evidence for strong and persistent X-ray variability, both on
short and long time-scales. On short time scales of $\gta$ 2 hours the 
source varies smoothly by 15$-$20\%; long term variations by a factor $\gta$ 2
could not be resolved in the relatively short exposures: we find the source
mostly in  a `low' state, in 2 out of the 11 observations in a `high state'.
X-ray flares seem to be recurrent with a time scale of $\sim$ 24 ksec
which, if interpreted as the Keplerian time scale in the disc, would
place the emission region just outside the last stable orbit. 
The X-ray spectrum of \pks ~can be well fitted by two Comptonization
components, one at moderate temperatures of kT $\sim$ 4.5 keV and
optical depths of $\tau \sim 2$, the other at high temperatures
(kT $\gta$ 50 keV) and low optical depths ($\tau \lta 1.0$). 
These parameters are, however, subject to large errors due to the 
inherent degeneracy of the Comptonization models. 
Flux variations of the source are caused by changes of the colder
  component only,
the hot component with parameters very similar to those of BLS1 galaxies,
stays constant.
All results fit nicely the picture that NLS1 galaxies are lower mass
objects, accreting close to the Eddington rate emitting X-rays from
a very active magnetically powered accretion disc corona.
 }
 
\titlerunning{X-ray variability of \pks}
\authorrunning{W. Brinkmann et al.}
\maketitle

\section{Introduction}  
Narrow-line Seyfert~1 galaxies are optically identified by their  emission 
line properties: the ratio [O III]/H$\beta$ is less than 3 and 
the FWHM H$\beta$
less than 2000${~\rm km~s^{-1}}$ (Osterbrock \& Pogge 1985, Goodrich 1989). 
Their spectra are further characterized by the presence of strong
 permitted  
Fe II, Ca II,  and O I $\lambda$ 8446 ionization lines (Persson 1988).
NLS1 are seldom radio loud (Ulvestad et al. 1995,
 Siebert et al. 1999, Grupe et al. 1999) and they 
are usually strong infrared emitters (Moran et al. 1996). 
In particular, NLS1  have been found to have extreme spectral  
and variability
properties in soft X-rays (Boller et al. 1996, Brandt \& Boller 1998, 
Boller et al. 2000), which might be related to an extreme value of a 
primary physical parameter,
originating from the vicinity of a super-massive black hole. 

PKS 0558-504 ($z=0.137, m_{\rm B}=14.97$) is one of the very few radio-loud 
NLS1 galaxies ($R_{\rm L}=f_{\rm 5 GHz}/f_{\rm B}\simeq 27$, Siebert et al.
1999). It was optically identified  on the basis of X-ray positions
from the High Energy Astronomy Observatory (HEAO-1, Remillard et al. 1986).
The Ginga observations (Remillard et al. 1991) showed an increase in the
X-ray flux by 67\% in 3 minutes, implying that the apparent
luminosity must be enhanced by relativistic beaming.
Further X-ray observations with different satellites have confirmed the
peculiar activity (e.g. strong variability, steep spectrum, high luminosity)
of this source. 
Gliozzi et al. (2000) reported long-term variability from different 
X-ray observations corresponding to  
luminosities in the 0.1$-$2.4 keV soft X-ray band
between 1.5$\times10^{45}$ \ergs and 5.4$\times10^{45}$\ergs
(for $H_0=50 {~\rm km ~s^{-1}~Mpc^{-1}}$, $q_0=0.5$ and isotropic 
emission).  The soft X-ray spectrum is rather steep ($\Gamma \sim 3.1$) and
the medium energy power laws are considerably flatter ($\Gamma \sim 2$).
Another peculiar property 
displayed by PKS 0558-504 is the unusually high X-ray to radio luminosity ratio
(Brinkmann et al. 1997). 

\begin{table*}
\tabcolsep1ex
\caption{\label{observations} XMM$-$Newton  PN /RGS observations of
          PKS 0558$-$504}
\begin{tabular}{c r c c c r c}
\noalign{\smallskip} \hline \noalign{\smallskip}
\multicolumn{1}{c}{Orbit} & \multicolumn{1}{c}{Observing date} &
\multicolumn{1}{c}{Instrument} & \multicolumn{1}{c}{Mode}      &
\multicolumn{1}{c}{Filter} &
\multicolumn{1}{c}{}Live time & \multicolumn{1}{c}{count rate} \\
\multicolumn{1}{c}{  } & \multicolumn{1}{c}{(UT)} &
\multicolumn{1}{c}{  } & \multicolumn{1}{c}{ } &
\multicolumn{1}{c}{sec} & \multicolumn{1}{c}{(ksec)} &
\multicolumn{1}{c}{count/s} \\

\noalign{\smallskip} \hline \noalign{\smallskip}
30&Feb 07, 2000: 11:16-13:19 &PN &FF &medium & 5.93 & 13.3 \\
32&Feb 10, 2000: 23:26-03:23 &PN &FF &thin2 & 11.49 & 18.2 \\
33&Feb 12, 2000: 23:13-04:41 &PN &FF &thin2 & 16.22 & 17.07\\	
 &Feb 13, 2000: 12:33-17:04 &PN &FF &thin1 & 13.34 & 12.41\\
 &Feb 13, 2000: 18:29-22:59 &PN &FF &thin2 & 13.30 & 13.74\\	
42&Mar 02, 2000: 18:15-21:42 & PN & FF & medium & 10.93 & 26.0\\
45&Mar 07, 2000: 20:49-03:31&PN&SW&thin1&16.88&26.9\\
84&May 24, 2000: 13:04-16:54&PN&FF&medium&10.68&14.5\\
&         17:21-21:08&PN&SW&medium&9.53&19.7\\
&         06:19-13:16&RGS&SP&-&25.00 &n/a \\
&         13:16-22:13&RGS&SP&-&32.21&n/a \\
153&Oct 10, 2000: 01:49-04:47&PN&SW&thin1&7.50&18.8\\
&         05:38-08:01&PN&FF&thin1&8.05&15.3\\
&         01:27-08:46&RGS&SP&-&26.41&n/a\\
283&Jun 26, 2001: 03:21-06:36&PN&FF&medium&10.59&12.2\\
&         02:37-06:43&RGS&SP&-&14.80& n/a \\
341&Oct 19, 2001: 12:41-15:51&PN&FF&medium&10.22&23.9\\
&         11:55-15:59&RGS&SP&-&14.61&n/a \\
\noalign{\smallskip}\hline
\end{tabular}
\medskip
\end{table*}

The six days long ROSAT All Sky Survey light curve of September 1990 and a
series of pointed HRI observations (April 19 - 25, 1998) showed
count rate variations by a factor of 2 in less than one day 
(Gliozzi et al. 2000).
With the most extreme value of the luminosity variations found during the
observations,
${\rm\Delta}L/{\rm\Delta}t=1.12\times 10^{42}~{\rm erg~s^{-2}}$,
the straightforward application for an estimate of the lower limit
of the radiative
efficiency of $\eta > 4.8\times 10^{-43}{\rm\Delta}L/{\rm\Delta}t$
 (Fabian 1979) led to $\eta > 0.54$, which exceeds
even the theoretical maximum for accretion onto a maximally rotating Kerr
black hole. However, spectral variations and the extrapolation of steep
power law spectra to low energies can lead  to uncertain luminosities,
as pointed out by Brandt et al. (1999).
~

Very dramatic flux variations by a factor of two in 33 min and, perhaps,
by 40\% on a time scale as short as 2 min were reported by Wang et al. (2001)
from ASCA observations on January 31, 2000 which leads to an even
higher value of the  radiative efficiency  of $\eta =0.9\pm0.2$.

PKS 0558$-$504 was observed repeatedly by XMM$-$Newton as a  calibration
 and performance variation (PV) target. The XMM observations benefit from
the high sensitivity of the instruments and the uninterrupted exposures
resulting from the eccentric satellite orbit.
O'Brien et al. (2001) published the spectral analysis of some preliminary 
data of the commissioning/CalPV phase. The 0.2$-$10 keV spectrum is dominated by a large
soft X-ray excess, which shows no evidence for absorption or
emission line features. A power law spectrum at higher energies (E $\geq$ 4 keV)
requires in the soft band additionally three black body components; the most 
physical explanation for the hot big blue bump is, however, Comptonization by a 
multiple temperature corona of an accretion disc.
 
Gliozzi et al. (2001) used data from the same observations for a study of the 
X-ray variability of the source.  The long term light curve shows persistent
variability with a tendency of the X-ray spectrum to harden when the count rate 
increases. The short term variability is characterized by smooth modulations
with a typical time scale of $\sim$ 2.2 hr and the most extreme count rate
variations found imply a radiative efficiency slightly higher than the 
theoretical maximum for accretion onto a Schwarzschild black hole.
 
In this paper we present an analysis of XMM PN and RGS data taken so far in
several observations between January 2000 and  October 2001.
The next section starts with  the observational details  and a temporal 
analysis of the data.
Sect. 3 deals with the spectral variability of 
PKS 0558-504. In Sect. 4 we discuss the results in the framework 
of disc-corona accretion models for Seyfert galaxies and the last
part contains a summary of the main conclusions.

\renewcommand{\topfraction}{0.99}
\begin{figure}[t]
\psfig{figure=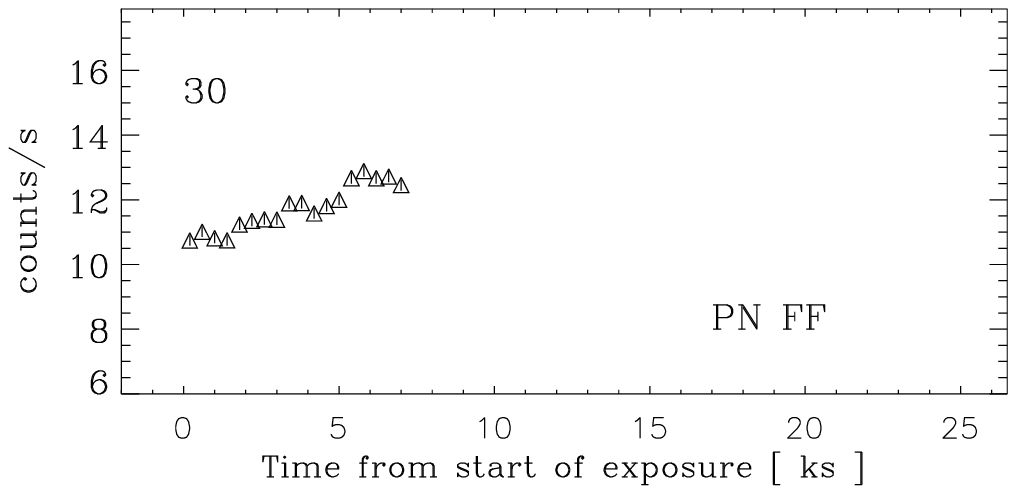,width=8.5cm,height=3.6cm,%
   bbllx=136pt,bblly=606pt,bburx=427pt,bbury=720pt,clip=}
\psfig{figure=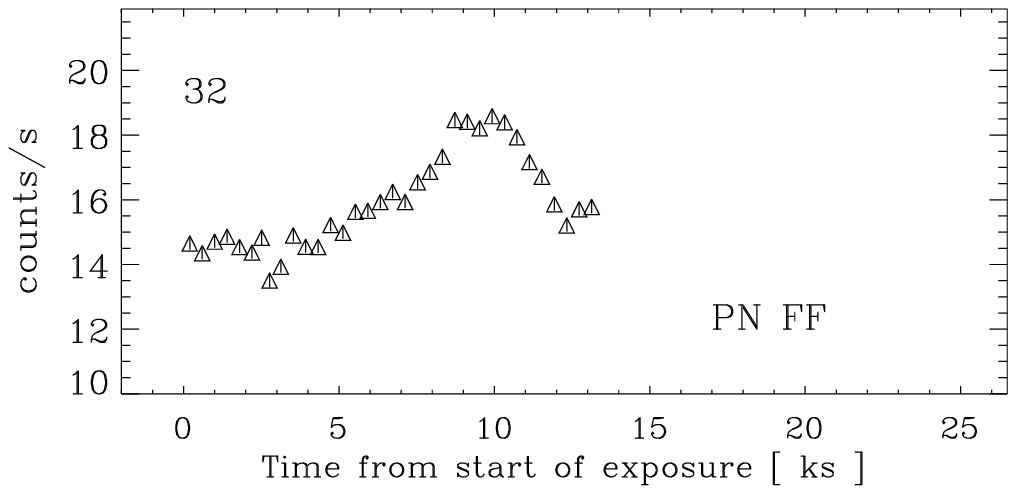,width=8.5cm,height=4.1cm,%
   bbllx=136pt,bblly=593pt,bburx=427pt,bbury=721pt,clip=}
\psfig{figure=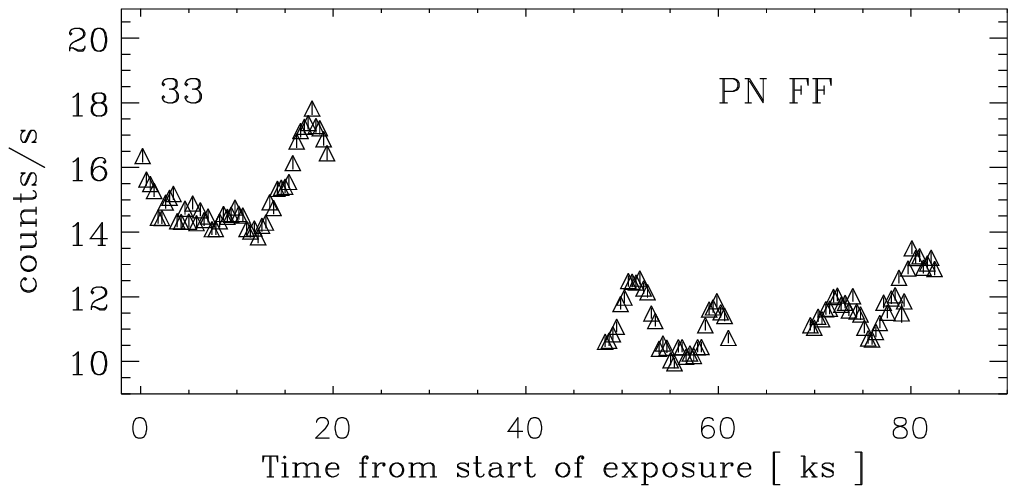,width=8.5cm,height=4.1cm,%
   bbllx=136pt,bblly=593pt,bburx=427pt,bbury=725pt,clip=}
\psfig{figure=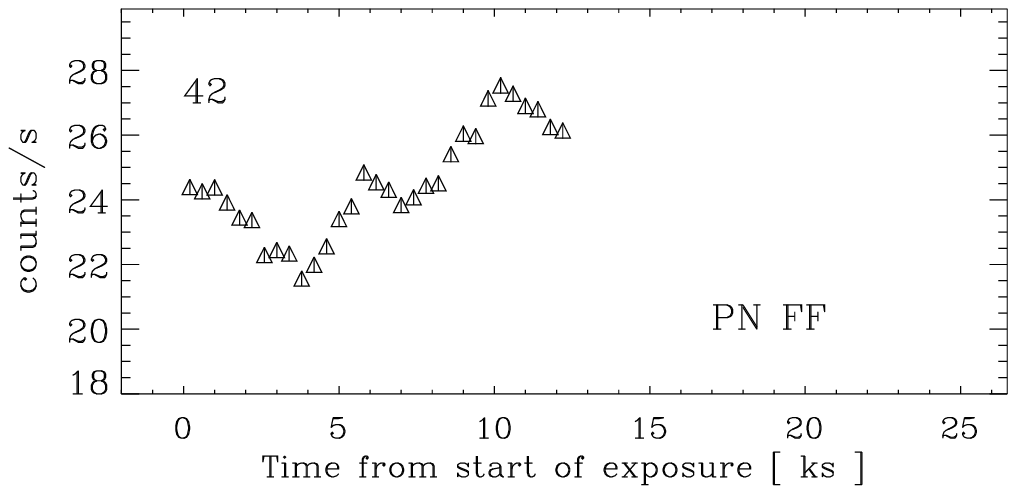,width=8.5cm,height=3.6cm,%
   bbllx=136pt,bblly=607pt,bburx=427pt,bbury=722pt,clip=}
\psfig{figure=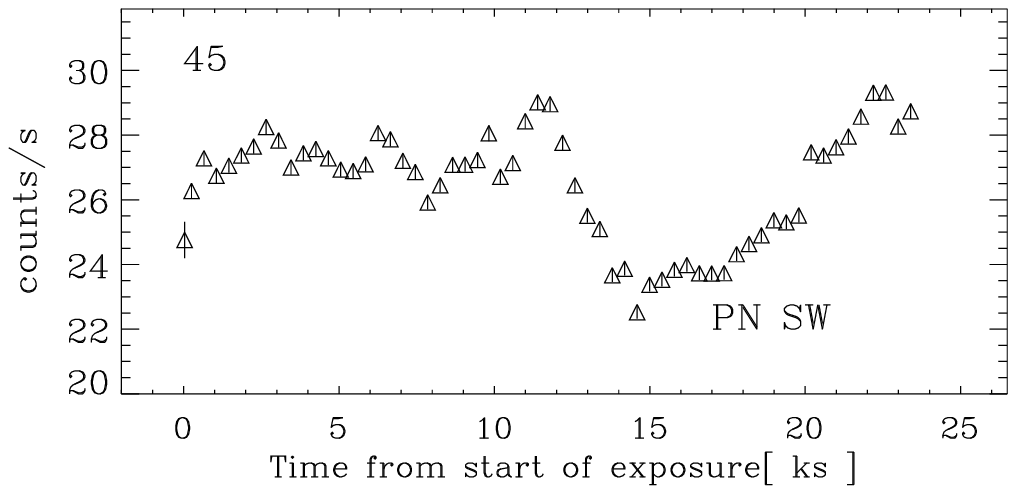,width=8.5cm,height=4.42cm,%
   bbllx=136pt,bblly=578pt,bburx=427pt,bbury=719pt,clip=}
\caption {Light curves of PKS 0558-504 for the early orbits without
RGS exposures.  Triangles represent the PN data, with error bars 
smaller than the symbols.
Note the longer time axis for orbit 33.}
\label{lightcurve1}
\end{figure}
\begin{figure}[t]
\psfig{figure=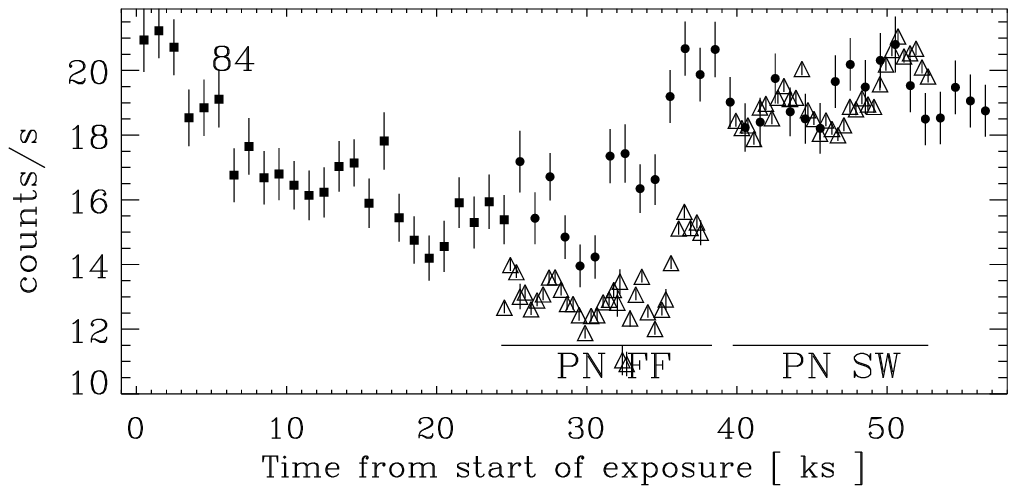,width=8.5cm,height=4.02cm,%
   bbllx=136pt,bblly=593pt,bburx=427pt,bbury=720pt,clip=}
\psfig{figure=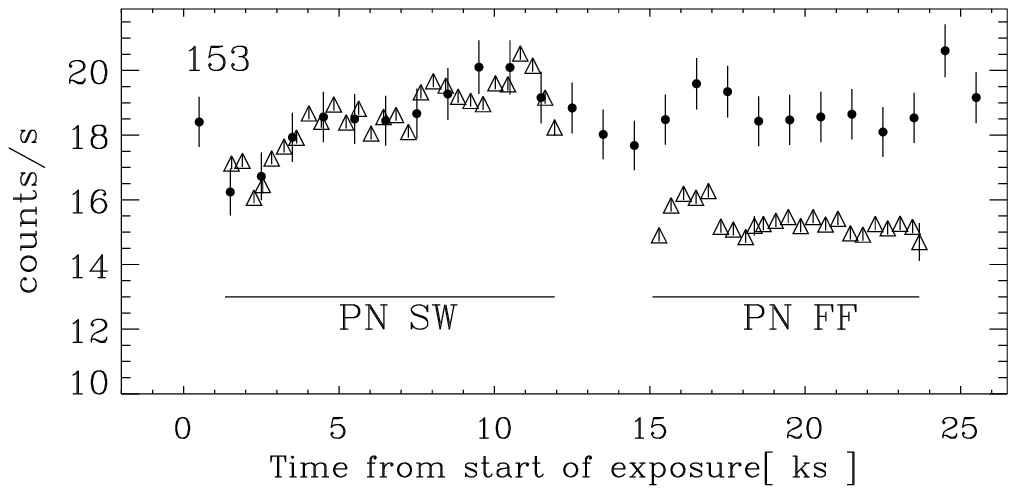,width=8.5cm,height=3.6cm,%
   bbllx=136pt,bblly=608pt,bburx=427pt,bbury=724pt,clip=}
\psfig{figure=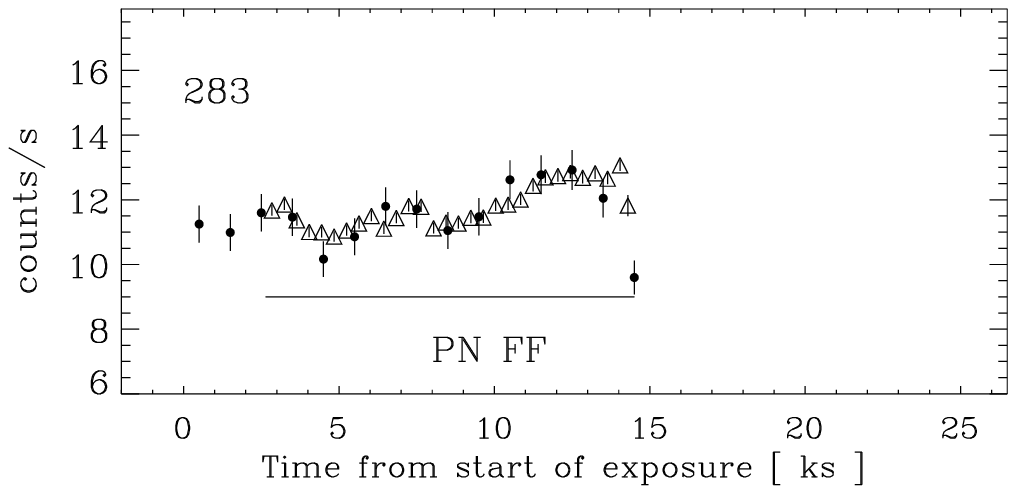,width=8.5cm,height=3.5cm,%
   bbllx=136pt,bblly=607pt,bburx=427pt,bbury=719pt,clip=}
\psfig{figure=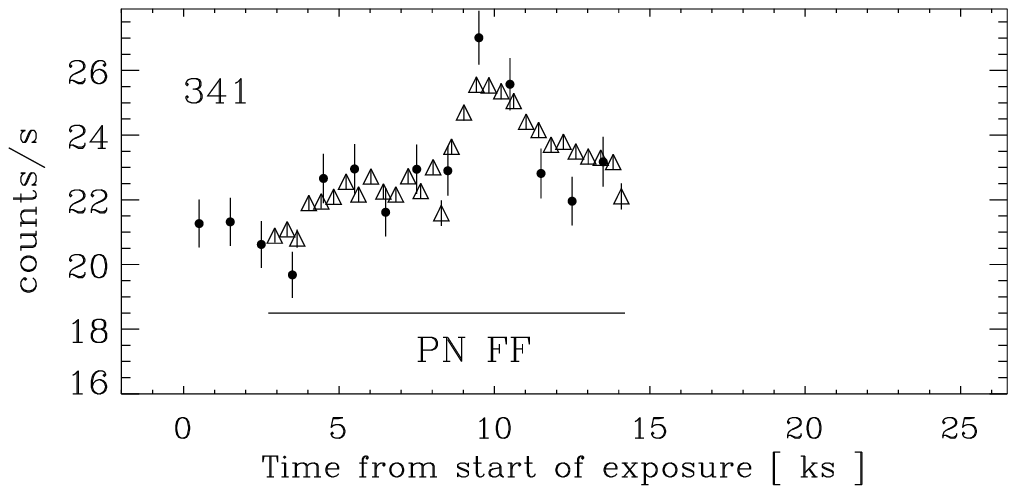,width=8.5cm,height=4.32cm,%
   bbllx=136pt,bblly=578pt,bburx=427pt,bbury=719pt,clip=}
\caption {Light curves of PKS 0558-504 for the later orbits with RGS
exposures. 
Triangles represent the PN data, with error bars smaller than the symbols.
The dots represent the RGS light curves rescaled to match the PN SW 
count rate of the corresponding exposure. 
For orbit 84 the filled squares represent the first
RGS exposure using both RGS detectors while for
the second RGS exposure (dots) only the RGS 2 was available.}
\label{lightcurve2}
\end{figure}
 
\section{Observations and temporal  analysis}

PKS 0558-504 was observed with  the instruments on board XMM$-$Newton
during 9 orbits between Jan 2000 and October 2001.
As the source was a calibration target  the
exposures of the EPIC instruments were always relatively short and 
different instrument modes were used.
For the following analyses we will rely only on data from the RGS 
instrument (den Herder et al. 2001) and from 
the PN camera (Str\"uder et al. 2001).
Table ~\ref{observations} lists the relevant data of the observations.
As count rate we quote the total number of counts on the chip,
divided by the live time.
The RGS were always operated in spectroscopy mode (SP), the PN camera 
in Small Window (SW)
or Full Frame (FF) mode, with either a thin or medium filter
(for details of the XMM
instruments see Ehle et al. 2001).

All PN data have been reprocessed using XMMSAS version 5.4.
We selected photons with Pattern $\leq$ 4
(i.e. singles and doubles) and quality flag = 0.
We calculated the  0.2$-$10 keV light curves for all observations by
extracting the photons from a
circular region centered on the source with a radius of 50\arcsec.
This extraction radius contains about 90 \% of the
source photons,
using the  encircled energy function given by Ghizzardi \& Molendi (2001).
We did not correct the light curves for the photon pile up which 
occurred in the central pixels  of the point spread function 
when the detector was operated in FF mode and the source was
in a bright state.
It amounts to about $\sim$ 3\% and  does not change the form of
the light curve significantly.    
The backgrounds were determined  with the same selection criteria
from source free regions on the same chip  and subtracted from the
source light curves, which were binned in 400 sec bins.
 
For four of the orbits the reflection gratings (RGS) provided
scientifically useful data simultaneously to the PN exposures.
The RGS data was processed using the task $rgsproc$ of the XMMSAS
version 5.4. The light curves were constructed using the task
$evselect$, selecting photons  in the 0.3$-$2.0 keV energy band 
that belonged both to the source spatial
region and the first order spectral region. The background light
curves were calculated from all photons 
belonging to the first order spectral region and to the spatial
region excluding the source.
As the detector area used for the background is
larger than the area of the source the background light curve
had to be rescaled before being subtracted. The background
subtracted source light curve was binned in 1000s bins to achieve a
reasonable signal to noise ratio while preserving the overall shape of
the curve.

The PN data is affected by the dead time of the detector that reduces
the effective exposure by a factor of 0.9994 in the FF mode and
0.7101 in the SW mode. There is also a loss of counts due to the
out-of-time events, the photons collected during the read out. This
last effect further reduces the collection efficiency by a factor 
of 0.9328 in the FF and 0.9884 in the SW mode, respectively.
 The PN light curves were thus corrected
for these two effects. The RGS light curves were scaled by a factor
of $\approxgt$ 12 in order to match them with the resulting PN SW count
rates of the corresponding orbits.
The plots in Fig.~ \ref{lightcurve1}
show the PN light curves of early orbits for which no RGS observations 
were performed.  In Fig.~ \ref{lightcurve2} we present the later 
PN observations with the scaled RGS light
curves overlaid so that the times match.
We used the same length for most of the  time axes and the same range of count
rates for a comparison of the  variability pattern.
Only  orbit 33 and  ~84
 had  considerably longer exposures.

Clearly visible is a deficit of counts in the FF mode  of $\sim 20$\% 
when the RGS light
curve is normalized to the SW mode data. This is an effect of
photon pile up and a redistribution of double events into  higher
order event patterns (Freyberg, private comm).
The correction factor to the RGS data is in this case of pattern 
migration of the order of
$\approxgt$ 15, but it is, as for the SW mode data, not the same for all 
orbits. It obviously does not only depend  on the count rate and 
spectral variations but seems to be influenced by the level of
background activity as well.
For the following  correlations of source properties with source
intensities we will use normalized count rates, obtained by 
detailed corrections for the above instrumental effects.  
 
On long time scales (different orbits) the light curves show 
that the source undergoes 
intensity variations by a factor of $\gta$ 2.
During the individual exposures we find substantial  variability 
with an amplitude of up to $\sim$ 15$ - $20\%, which is not correlated 
with the average count rate of the observation.
We never detected the large intensity variations in an individual
exposure,  thus 
the time scale must be longer than typically one orbit.
Fig. \ref{figure:HRs} (see sect. 2.1) further suggests that the 
source is mainly found in a low state, with only relatively rare
high states. 
The variances in the hard and soft band light curves are very 
similar; only in the highest intensity states the hard band light curves 
show stronger variations than the soft light curves.
 
\begin{figure}
\psfig{figure=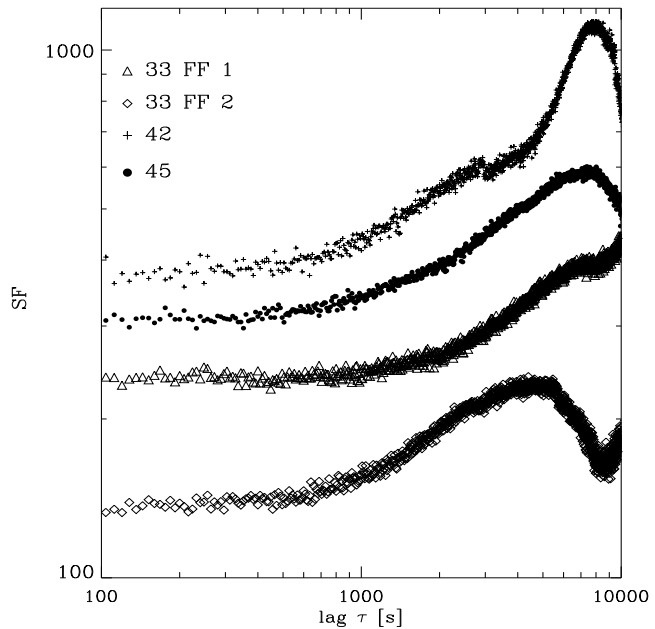,width=8.5cm,height=8.5cm,angle=0,%
 bbllx=125pt,bblly=380pt,bburx=440pt,bbury=725pt,clip=}
\caption{Structure functions of the light curves of several orbits.
 The graphs are displaced arbitrarily in y-direction for 
clarity.}
\label{figure:SFs}
\end{figure}
 
The relatively smooth  and nearly linear intensity changes 
have a time scale of typically \gta 2 
hours as can be seen directly from the light curves (Figs. 1 and 2).
However, this is not a strict, well defined time scale as illustrated
by the structure functions (Simonetti et al. 1985)
 of some individual orbits (Fig. \ref{figure:SFs}). 
As already noted by Gliozzi et al. (2001) this time scale merely seems to 
reflect the typical rise/decay time 
scales in the light curves  with a relatively narrow variance. 
But the light curves show as well that even the longest RGS exposure
in orbit 84 is too short to cover all relevant time scales 
of the source.
 
The temporal structure of the intensity variations during different orbits show
some intriguing similarities in shape and even in the amplitudes of 
the flux variations, independent of the average flux level.
A temporal analysis  of the whole data set,
using orbit 32 as a template, shows a `repetition rate'
 of that signal of $\sim$ 23850 sec.
The search was simply done by aligning the count rate peak of 
orbit 32 with the individual peaks of the other light curves.
If the signal were strictly periodic we would expect this flare in
 8 of the orbits but we find only 6 at the expected position. 
There are two exposures,
orbits 153 and 283, when the source is in a low state, where no flare is
found at the expected time. 
The statistical significance for this `quasi-periodic' repetition
 would be rather high
for purely random fluctuations, but lower for other models like shot-noise or 
self-organized criticality in the accretion disc (Mineshige et al. 1994).
A rigorous statistical analysis to determine the significance of
the existence of that signal, like the Monte Carlo technique developed
by Uttley et al. (2002)  turned out to be impossible, due to the
underlying red noise process and, in particular, due to the extreme 
sparsity of the data, covering observations over more than two years.
Longer observations are clearly required for definite results.
  
The most dramatic flux variations occurs near the middle of orbit 341 
where the count rate changes by $\sim$ 4.3 counts/s in $\sim$ 1128~s.
With the spectral parameters deduced below, assuming a Friedman 
cosmology with H$_o$ = 75 km\,s$^{-1}$Mpc$^{-1}$ and q$_o$=0.5,  we
obtain a ${\rm\Delta}L_{0.2-10{\rm keV}}/{\rm\Delta}t $
$ \sim 4.5\times 10^{41}$ erg\,s$^{-2}$. 
This leads to an upper limit of the radiative 
efficiency (Fabian 1979) of $\eta \sim 0.23$ which is similar to that
found previously from ROSAT (Gliozzi et al. 2000) and XMM observations
(O'Brien et al. 2001). These efficiency values are much more moderate
than those claimed for the two flares detected by Ginga (Remillard et al.
1991) and Wang et al. (2001).
 
\subsection{Hardness Ratios}

We computed hardness ratios using 1000~sec binned soft and hard band
light curves, and the expression, $HR=[4.0-10$~keV$]/[0.2-1.0$~keV].
The count rates in these energy bands appear to represent the two relatively
independent spectral components of the source, as can be seen in sect. 4. 
By plotting the $HR$ as a function of the total count rate (0.2$-$10~keV),
 we are
able to investigate whether the $HR$ variations are correlated with the
source flux. 

The total count rates were corrected for the above mentioned 
dead-time effects as well as for pattern migration and pile up.
From spectral simulations with typical source parameters (see sect. 3)
we found that the hardness ratios  change by about 0.01  comparing 
the thin filter observations with the medium filter data. 

\begin{figure}
\psfig{figure=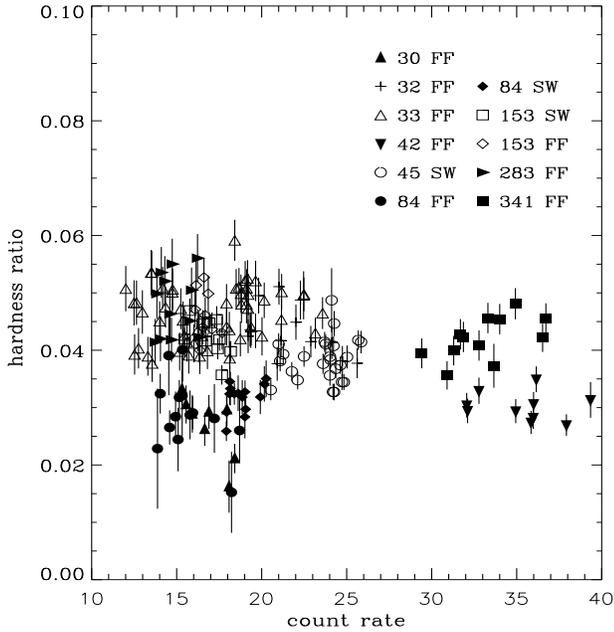,width=8.5cm,height=8.5cm,angle=0,%
 bbllx=115pt,bblly=344pt,bburx=436pt,bbury=723pt,clip=}
\caption{Hardness ratios as function of the total count rate
 for different orbits. Filled symbols denote observations with
 medium filter, open symbols with thin filters.}
\label{figure:HRs}
\end{figure}
 
In Fig. \ref{figure:HRs} we plot the hardness ratios as 
function of the corrected 0.2$-$10 keV count rate for the different orbits.
Filled symbols denote observations with
 medium filter, open symbols with thin filter.
The count rates and hardness ratios are corrected to medium filter observations.
Some data points with large errors are affected by flares in the background.
We do not find a simple correlation between hardness ration and
count rate as claimed by Gliozzi et al. (2001) from their restricted
data base. 
The data suggest that the source can be predominantly found in 
 a lower intensity state, differing by a factor of two in count rate
from the high states. This reflects the skew distribution of 
count rates found in other NLS1  as well. 
Whether the gap between these two states is real or 
a selection effect due to the limited observational coverage, can only be
 decided in future observations.
The hardness ratios show a large dispersion on long
time scales, i.e. for different orbits.
On the shorter time scales of a typical observation the hardness ratios
in the low state appear to be nearly constant.
There is  a general slight tendency for the hardness ratios 
to get smaller with increasing count rate.
In the high state orbit 341 there are strong indications of a hardening 
of the spectrum when the source brightens.

\section{Spectral analysis}
 
NLS1 galaxies are generally characterized by very steep spectra in the soft 
energy band (Boller et al. 1996).
From \ros observations of \pks ~ Gliozzi et al. (2000) find a steep 
power law with
$\Gamma \sim 3$ in the 0.1$-$2.4 keV energy range; in the \asca ~0.6$-$10 keV
band Vaughan et al. (1999) obtain  $\Gamma = 2.26 \pm0.03$.
The XMM$-$Newton data (O'Brien et al. 2001) clearly show a strong soft excess 
below $\sim $2 keV over a harder power law at higher energies.
 
The PN data with their outstanding signal to noise ratio and their wide 
bandpass are ideally suited for a detailed spectral study of the source.
For the spectral analysis we used the latest available response matrices
(version 6.3) issued in December 2002.
We extracted single and double events with quality flag = 0
from a rectangular region of 30$\times$30 RAW pixels around the source
position.
The region includes $\sim$ 90\% of the source photons 
but avoids the gap between the detector chips.
The background was taken on the same chip at distances as close to the source
position as possible, avoiding contamination.
With a count rate of $\gta$ 20 counts~s$^{-1}$ in the high state the 
PN detector,
operated in Full Window mode, showed strong indications of pile-up, clearly
apparent from the XMMSAS task $epatplot$.
We therefore discarded photons from the innermost (typically $2\times3$)
 RAW pixels
at the core of the point spread function from the spectral analysis.

\begin{figure}
\psfig{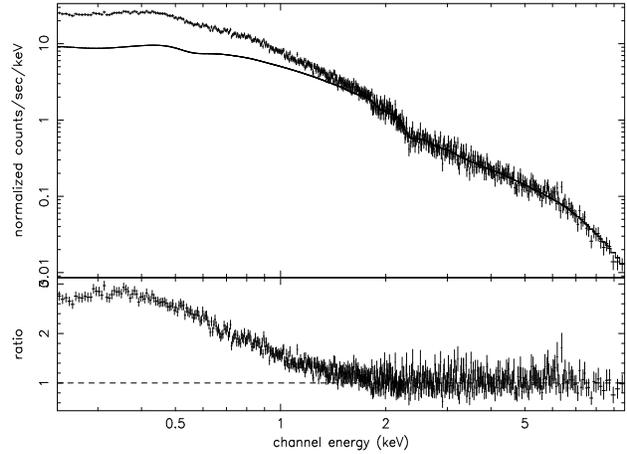}
\caption[]{Power law fit with galactic absorption to the PN data for \pks ~in the
 $2 - 10$ ~keV energy range; the fitted model is extrapolated to lower energies.
 The lower panel gives the ratio between data and model.}
\label{figure:powl}
\end{figure}

In Fig. \ref{figure:powl} we show the power law fit 
in the 2$-$10 keV energy range to the data of orbit 153  ($\Gamma = 2.13\pm0.03;
\chi_{\rm red}^2 = 0.90 / 303$ d.o.f with a galactic N$_H = 4.4\times 10^{20}$ 
cm$^{-2}$).  The fitted model is extrapolated to lower energies and the
ratio between data and model, given in the lower panel, clearly demonstrates
the presence of a large soft X-ray excess over the hard power law. 
The spectrum of \pks ~ appears to be  a carbon copy
of that of  the  NLS1 galaxy PG~0844+349 (Brinkmann et al. 2003) 
even with respect to the `big blue bump' seen in both objects (O`Brien et al. 2001)
and it is very similar to that of 1H 0419-577 (Page et al. 2002) and 
the other NLS1 galaxy Mrk 896 (Page et al. 2003) studied with XMM-Newton.

The upper limits given for an iron line are rather low (O'Brien et al. 2001)
and  the soft band  spectral excess is far too broad to be 
fitted by a single black body component.
A multiple blackbody (in the soft band) plus a power law at higher energies 
provides acceptable fits to the data, however, the physical nature of these 
different components remains obscure. 
 While a model with two Comptonization components gives an
accurate description of the spectra,  two power law models require absorption
 in excess of the galactic value and yield slightly worse fits.
 For example, for the above 
mentioned orbit 153 (see as well Tab. 2) we obtain an 
N$_H = 7.79\times10^{20}$ cm$^{-2}$, $\Gamma_{\rm soft}=3.28\pm0.08$,
$\Gamma_{\rm hard}=1.62\pm0.10$ with a
$\chi_{\rm red}^2 = 1.163 / 578$ d.o.f., values, which are representative 
for the other orbits as well.
 
\begin{figure}
\psfig{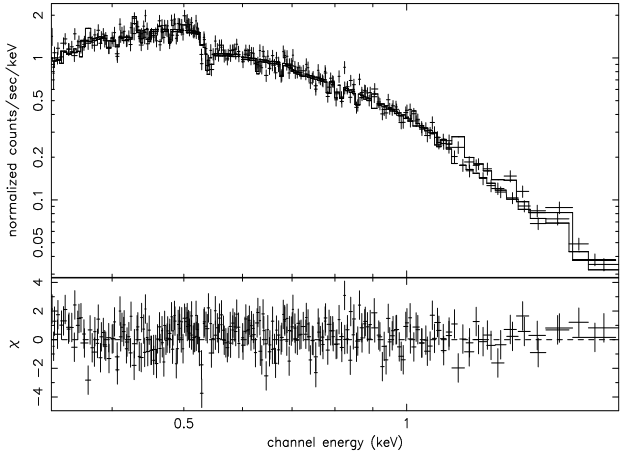}
\caption[]{CompTT+ power law  fit  (without edge) with galactic absorption 
 to the RGS1 data of orbit 84.
 The lower panel gives the $\Delta \chi^2$ per channel.}
\label{figure:rgs84}
\end{figure}

The relatively long RGS  observation of orbit 84 provides a sufficient number of
photons for an accurate fit.
We have reprocessed the RGS data using  XMMSAS version 5.3.3 and 
the RGS response matrices were created with the SAS package
$rgsrmfgen$. The spectral data were binned 
to contain at least 30 photons per energy channel.  
We fitted the RGS data with a single $compTT$ Comptonization model,
available in $Xspec$ (Titarchuk 1994), assuming
galactic absorption plus an additional high energy power law (see sect. 3.1). 
The fit is acceptable ($\chi^2_{\rm red}= 1.01/1162$ d.o.f) and
yielded a temperature of the soft photons of kT$_0 = 63.9 \pm 5.0$ eV
 and an optical depth of $\tau = 2.47\pm 0.25$. 
The temperature of the hot Comptonizing electrons  was fixed at 
kT$_1 = 4.5$ keV to reduce the error bars (see the discussion below). 
An additional absorption edge, required only for one of the two 
observation intervals (E$_{\rm edge} = 1.14 \pm 0.02$ keV, 
$\tau = 0.27\pm0.06$) improves the fit  significantly 
($\chi^2_{\rm red}= 0.989/1158$ d.o.f) according to an F-test.
The residuals  shown in Fig. \ref{figure:rgs84} are, however,
 smaller than the
systematic differences between the two RGS detectors and no other 
strong absorption or emission features are observed.
   
We tried to determine the possible contribution of an iron line (neutral or
ionized) to the spectra.
To improve the signal to noise ratio  we accumulated the data
from different
observation modes and filters and fitted first a power law and then
a power law with a line to the combined SW and FF - mode 4$-$9 keV data.
 As the accumulated spectra for the
different modes covered different intensity states of the source, the
normalizations of the spectra were different for the SW and FF mode data.
For a neutral iron line we obtained  90\% upper limits for the  equivalent
 widths of EW $<$ 25~eV (SW) and
$<$ 35 eV for the FF - mode data; for an ionized
line we obtained EW $<$ 54 eV (SW) and $<$ 72 eV for the FF-mode data, using
a line width of $\sigma = 0.1$ keV for the line. For narrow lines ($\sigma = 0.01$ keV)
the corresponding limits are nearly a factor of 2 lower.
In all fits 
the inclusion of the lines did not improve the fits. The single power law
model yielded already $\chi^2_{\rm red} = 0.927$, thus the data do 
not require the presence of any iron emission line.

\begin{table*}
\small
\tabcolsep1ex
\caption{\label{fits} Results from spectral fits of two$-$component Comptonization 
models  assuming fixed  \hfill \break
\hspace*{1.5cm}Galactic N$_{\mathrm H} = 4.4\times10^{20}$cm$^{-2}$, z=0.137}

\begin{tabular}{llccclclcc}
\noalign{\smallskip} \hline \noalign{\smallskip}
\multicolumn{1}{c}{Orbit,} & \multicolumn{1}{c}{kT$_o$} &
\multicolumn{1}{c}{kT$_{1}$} & \multicolumn{1}{c}{$\tau_{1}$} &
\multicolumn{1}{c}{norm$_{1}$} & \multicolumn{1}{c}{kT$_{2}$}
& \multicolumn{1}{c}{$\tau_{2}$} 
 & \multicolumn{1}{c}{norm$_{2}$} & \multicolumn{1}{c}{$\chi ^2_{\rm red}$} & 
\multicolumn{1}{c}{dof} \\
\multicolumn{1}{c}{mode} & \multicolumn{1}{c}{(eV)} & \multicolumn{1}{c}{ (keV)}
&\multicolumn{1}{c}{ } & \multicolumn{1}{c}{} & \multicolumn{1}{c}{(keV) } &
\multicolumn{1}{c}{} & \multicolumn{1}{c}{} & \multicolumn{1}{c}{} &
\multicolumn{1}{c}{}\\
\noalign{\smallskip} \hline \noalign{\smallskip}
~30 FF     & 77.1 $^{+10.2}_{-10.2}$& 4.5 $^{+166.8}_{-4.5}$& 1.9 $^{+60.0}_{-1.9
}$& 1.30E-2& 47.3 $^{+2614}_{-47.3}$& 0.72 $^{+51.77}_{-0.72}$& 1.90E-4& 1.098& 464\\
~32 FF     & 87.4 $^{+5.6}_{-5.6}$& 4.6 $^{+74.8}_{-4.6}$& 2.4 $^{+29.7}_{-2.4}$&
 1.21E-2& 61.8 $^{+3929}_{-61.8}$& 0.93 $^{+70.16}_{-0.93}$& 0.98E-4& 1.026& 716\\
~33 FF (1) & 84.8 $^{+5.1}_{-5.1}$& 4.4 $^{+68.4}_{-4.4}$& 2.3 $^{+27.5}_{-2.3}$&
 1.10E-2& 52.4 $^{+1960}_{-52.4}$& 0.77 $^{+36.66}_{-0.77}$& 1.97E-4& 1.039& 823\\
~33 FF (2) & 86.9 $^{+6.9}_{-6.9}$& 5.3 $^{+119.6}_{-5.3}$& 1.9 $^{+37.7}_{-1.9}$
& 0.71E-2& 60.2 $^{+2869}_{-60.2}$& 0.73 $^{+44.73}_{-0.73}$& 0.96E-4& 1.089& 639\\
~33 FF (3) & 82.8 $^{+7.4}_{-7.4}$& 5.3 $^{+138.3}_{-5.3}$& 1.8 $^{+42.0}_{-1.8}$
& 0.86E-2& 61.2 $^{+2144}_{-61.2}$& 0.63 $^{+29.34}_{-0.63}$& 1.34E-4& 1.088& 663\\
~42 FF     & 86.7 $^{+5.3}_{-5.3}$& 4.5 $^{+68.38}_{-4.5}$& 2.4 $^{+27.8}_{-2.4}$
& 2.43E-2& 51.4 $^{+2502}_{-51.4}$& 0.60 $^{+40.15}_{-0.60}$& 3.38E-4& 1.029& 757\\ 
~45 SW     & 74.7 $^{+6.2}_{-6.2}$& 4.8 $^{+103.8}_{-4.8}$& 1.9 $^{+35.7}_{-1.9}$
& 3.37E-2& 49.9 $^{+2282}_{-49.9}$& 1.08 $^{+55.46}_{-1.08}$& 2.74E-4& 1.002& 909\\
~84 FF     & 78.8 $^{+17.4}_{-17.4}$& 4.5 $^{+241.3}_{-4.5}$& 1.7 $^{+84.0}_{-1.7
}$& 1.59E-2& 49.9 $^{+2782}_{-49.9}$& 0.59 $^{+45.27}_{-0.59}$& 2.92E-4& 0.957& 415\\
~84 SW     & 73.4 $^{+11.8}_{-11.8}$& 4.6 $^{+187.5}_{-4.6}$& 1.8 $^{+65.9}_{-1.8
}$& 2.73E-2& 52.8 $^{+2659}_{-52.8}$& 0.72 $^{+46.16}_{-0.72}$& 3.05E-4& 1.029& 485\\
153 FF    & 87.3 $^{+8.8}_{-8.8}$& 4.5 $^{+112.3}_{-4.5}$& 2.0 $^{+42.5}_{-2.0}$
& 0.98E-2& 49.8 $^{+2440}_{-49.8}$& 0.72 $^{+46.25}_{-0.72}$& 1.98E-4& 1.067& 576\\
153 SW    & 77.0 $^{+7.8}_{-7.8}$& 4.7 $^{+119.7}_{-4.7}$& 1.9 $^{+42.1}_{-1.9}$
& 2.18E-2& 50.5 $^{+1974}_{-50.5}$& 0.79 $^{+38.96}_{-0.79}$& 3.00E-4& 0.974& 656\\
283 FF    & 87.8 $^{+10.1}_{-10.1}$& 5.2 $^{+184.5}_{-5.2}$& 1.7 $^{+55.3}_{-1.7
}$& 0.68E-2& 40.0 $^{+1135}_{-40.0}$& 0.83 $^{+29.74}_{-0.83}$& 3.97E-4& 1.119& 695\\
341 FF    & 93.1 $^{+6.5}_{-6.5}$& 5.0 $^{+122.8}_{-5.0}$& 2.1 $^{+42.6}_{-2.1}$
& 0.80E-2& 46.3 $^{+1073}_{-46.3}$& 0.51 $^{+13.86}_{-0.51}$& 6.32E-4& 1.108& 792\\

\noalign{\smallskip}\hline

\end{tabular}
\medskip

NOTE: For orbits $<$ 45 the lower threshold for the event amplitudes 
 was different from the later setting.  \hfill \break \hspace*{1.1cm} No 
 extra response matrices exist for these early  observations, but the 
 spectral fits seem not to be \hfill \break \hspace*{1.1cm} 
 affected noticeably.
\end{table*}

The featureless spectrum indicates that we are seeing the
bare continuum disk emission from the quasar: thus disk Comptonization models,
where the X-rays are produced via inverse Compton emission in a hot
corona embedding a cooler accretion disk
(e.g. Haardt \& Maraschi 1993, Pounds et al. 1995),
 provide a satisfactory physical explanation to the data.
  
In a first step we fitted the combined PN data of each orbit, regardless of flux
variations during the observation with the sum of two $compTT$ components
with absorption fixed at the galactic value.
This Comptonization model describes the up-scattering of soft photons by hot, 
thermal electrons in a corona above the accretion disc.
The temperatures of the soft photons were assumed to be the same for both 
components.
In Tab. \ref{fits} we present the results of the fits for the 
different orbits and
observation modes. The temperatures of the soft photons, given in column 1 
are always around 80~eV, with relatively small errors.
The temperatures of the
Comptonizing electrons of the first component turned out to be k\,T$_1$ = 
4.5$-$5~keV, the optical depths around $\tau_1 \sim 2$, 
both with rather large errors.
The second component is characterized by rather high temperatures,
k\,T$_2$ \gta 50 keV, but optical depths $\tau_2$ \lta 1.0, 
and both parameters  have very large errors.
 
The physical parameters in the Comptonization models are only poorly constrained
because the spectral shape in the X-ray range is mainly determined by the 
combination of the  temperature kT/mc$^2$ and the optical depth $\tau$ of the
scattering electrons; the cutoff energy is essentially related to kT/mc$^2$.
For Comptonizing electrons as hot as found from the fits, the determining
spectral characteristics (high energy cut-off) is at energies far above the
energy range of the XMM instruments and only the spectral index $\alpha$ 
of the non-relativistic, low-energy part of the spectrum can 
be obtained (Titarchuk \& Lyubarskij 1995). 
Variations of the spectral shape with intensity, 
time lags between the fluxes in different energy bands, or energy 
resolved power density spectra as
additional diagnostic tools (e.g. Kazanas et al. 1997) might provide
insight into the way the corona is heated and cooled. 
  
\subsection{High energy power law fits}
  
While the cooler Comptonization model nicely matches the PN energy band
only the power law like, low energy part of the hot Comptonizing component 
falls into the energy band $\leq$ 10 keV. We therefore fitted the 
data with a combination of a $compTT$ plus a power law model. The temperature
of the Comptonizing photons was fixed at k\,T = 4.5~ keV and the absorption
to the galactic value. 

Table \ref{pofits} summarizes the results of these fits. The parameters of the
Comptonization model are rather well determined, the errors of the 
normalization  are typically $\la$ 6\%. The slopes of the high energy power 
laws are reasonably accurate, the normalizations of the power laws show,
however, rather large errors which can amount to $\la$ 30\% during the high
states, when the power law contribution to the total spectrum is small.

\begin{figure}
\psfig{figure=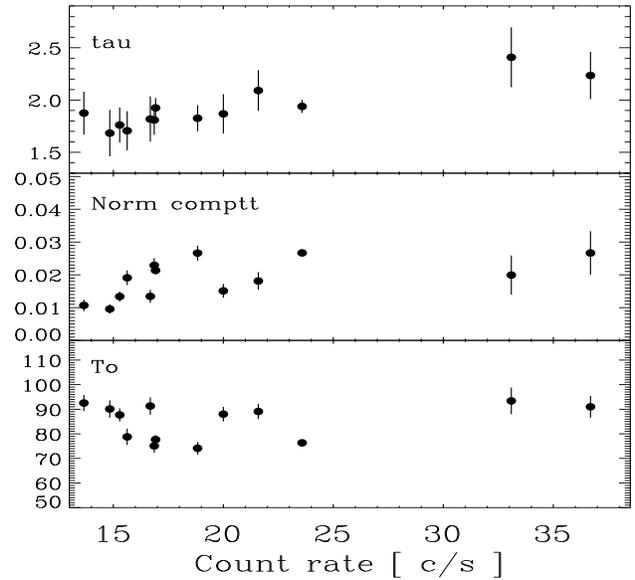,height=8.0truecm,width=8.3truecm,angle=0,%
 bbllx=143pt,bblly=368pt,bburx=428pt,bbury=721pt,clip=}
\caption[]{Parameters of the compTT component of a compTT + power law fit as 
 a function count rate. } 
\label{figure:comppar}
\end{figure}

\begin{figure}
\psfig{figure=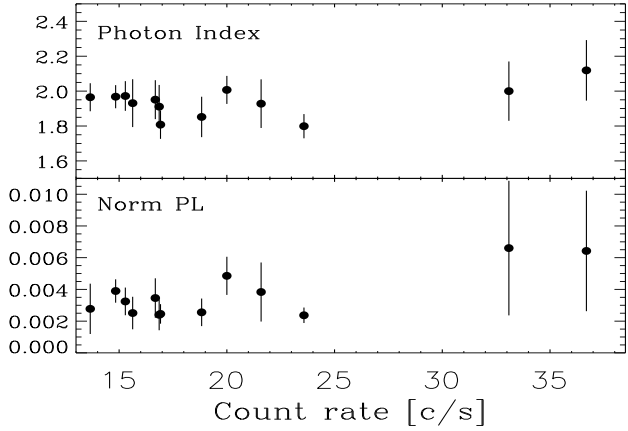,height=5.7truecm,width=8.3truecm,angle=0,%
 bbllx=137pt,bblly=185pt,bburx=428pt,bbury=426pt,clip=}
\caption[]{Parameters of the power law component of a compTT + power law fit as a function count rate.} 
\label{figure:popar}
\end{figure}

In Fig. \ref{figure:comppar} we show the dependencies of fitted parameters 
 of the Comptonization component on the 
corresponding count rates of the observation interval. The 
temperature T$_0$ of the  soft input photons (bottom panel) remains around
80 eV, with no clear correlation with the intensity state of the source
(a linear fit yielded a slope of $\beta = (0.290\pm0.299)\times 10^{-3}$).
 The optical depth shows a linear correlation with count rate 
with a slope of $\beta = (0.272\pm0.040)\times10^{-1}$.
 The parameters of the power law components of the same fits are shown in 
Fig. \ref{figure:popar}.
 The spectral index remains nearly constant  (slope of linear
fit $\beta=(0.593\pm0.324)\times10^{-2}$) at a value of 2.0.
Note, that in most plots the intrinsic variance of the data is
quite high.

The normalization of the power law changes only marginally and  
quite generally, the errors of the fitted parameters are larger
when the source is bright and thus the relative contribution of the 
power law component smaller. 
Changes of the flux are thus mainly  caused by an increase of
the optical depth of the scattering electrons combined with
an increase of the normalization of the Comptonization component.

We repeated the fits, leaving the temperature of the scattering electrons
free but fixed the optical depth to $\tau = 1.8.$
The results are very similar to those of Fig. \ref{figure:comppar}.
The flux variations are in that case correlated  linearly only
 with changes of the 
temperature  of the scattering electrons over the range 
3.9 keV \lta  kT \lta 7.4 keV. 


\section{Discussion}
\subsection{Spectral properties}
The spectrum of \pks , like that of other NLS1 galaxies, can be fitted well
by a combination of two compTT Comptonization models: 
one with low ($\sim$ 4.5 keV) temperature and moderate optical depths
 ($\tau \sim 2.0$) and a second, high
temperature component with kT $\gta$  50 keV and low scattering depths
 ($\tau \lta$ 0.7).
A determination of exact parameters  of the scatterer is not possible:
 from the high kT
component only the low-energy power law can be seen in the XMM energy band
and the measured slope is merely an indicator for a wide range of possible
(kT,$\tau$) combinations (Titarchuk \& Lyubarskij 1995).
 The low temperature component shows in principle measurable  
characteristic changes of the spectral slope inside the PN energy band;
 however, the superposition of the two components and the position 
of the spectral break
at rather high energies ($\gta 5$ keV) with low photon statistics
leaves considerable uncertainties in the parameter determination as well.
Thus, this low-temperature component could be quite well replaced with 
a considerably hotter component with much smaller optical depth; the 
physical implications of this  (less favored) combination will be 
discussed below. 
 
Nevertheless, the above envisaged two components,  low kT and high $\tau$ plus
high kT and low $\tau$  represent  an appealing picture for the X-ray emission
of NLS1 galaxies.
The high temperature component is very similar to that describing the emission 
of broad line Seyfert~1 (BLS1) galaxies. 
For those BeppoSax measurements (Petrucci et al. 2001)
gave temperatures in excess of 150 keV and optical depths around 
$\tau \sim $ 0.2. As the PDS detector covers the high energy band 
where spectral curvatures occur,
the errors on temperature and optical depth are very small;
the largest uncertainties seem to be related to the exact spectral shapes
predicted by the different Comptonization models.
 
\begin{table*}
\small
\tabcolsep1ex
\caption{\label{pofits} Results from spectral fitting  of a $compTT$ plus a
 hard power law model, assuming  \hfill \break \hspace*{1.5cm} fixed galactic
N$_{\mathrm H} = 4.4\times10^{20}$cm$^{-2}$, z=0.137}
\begin{tabular}{lcccclclcc}
\noalign{\smallskip} \hline \noalign{\smallskip}
\multicolumn{1}{c}{Orbit,} & \multicolumn{1}{c}{kT$_o$} &
\multicolumn{1}{c}{kT} & \multicolumn{1}{c}{$\tau $} &
\multicolumn{1}{c}{norm$_{compTT}$} & \multicolumn{1}{c}{$\Gamma$}
& \multicolumn{1}{c}{norm$_{po}$}
 & \multicolumn{1}{c}{$\chi ^2_{red}$} & \multicolumn{1}{c}{dof} \\

\multicolumn{1}{c}{mode} & \multicolumn{1}{c}{(eV)} & \multicolumn{1}{c}{ (keV)
}
&\multicolumn{1}{c}{ } & \multicolumn{1}{c}{} & \multicolumn{1}{c}{} &
\multicolumn{1}{c}{} & \multicolumn{1}{c}{} & \multicolumn{1}{c}{}\\
\noalign{\smallskip} \hline \noalign{\smallskip}
~30FF 	 & 75.1 $\pm 2.8$ & 4.5 & 1.81 $\pm 0.14$ & 2.29E-02 & 1.91 $\pm 0.12$ & 2.40E-03 & 1.104 & 466\\
~32FF 	 & 89.1 $\pm 3.1$ & 4.5 & 2.09 $\pm 0.19$ & 1.81E-02 & 1.93 $\pm 0.14$ & 3.84E-03 & 1.033 & 718\\ 
~33FF (1)& 88.0 $\pm 3.0$ & 4.5 & 1.87 $\pm 0.19$ & 1.51E-02 & 2.01 $\pm 0.08$ & 4.86E-03 & 1.045 & 825\\
~33FF (2)& 92.5 $\pm 3.3$ & 4.5 & 1.88 $\pm 0.20$ & 1.07E-02 & 1.97 $\pm 0.08$ & 2.78E-03 & 1.082 & 641\\
~33FF (3)& 87.7 $\pm 2.7$ & 4.5 & 1.76 $\pm 0.17$ & 1.34E-02 & 1.97 $\pm 0.09$ & 3.24E-03 & 1.076 & 665\\
~42FF 	 & 91.0 $\pm 4.4$ & 4.5 & 2.24 $\pm 0.23$ & 2.67E-02 & 2.12 $\pm 0.17$ & 6.43E-03 & 1.025 & 759\\
~45SW 	 & 76.3 $\pm 1.2$ & 4.5 & 1.94 $\pm 0.06$ & 2.67E-02 & 1.80 $\pm 0.07$ & 2.37E-03 & 1.040 & 856\\
~84FF 	 & 78.8 $\pm 3.4$ & 4.5 & 1.71 $\pm 0.19$ & 1.91E-02 & 1.93 $\pm 0.14$ & 2.51E-03 & 0.953 & 417 \\
~84SW 	 & 74.1 $\pm 2.6$ & 4.5 & 1.83 $\pm 0.13$ & 2.66E-02 & 1.85 $\pm 0.12$ & 2.56E-03 & 1.025 & 487\\
153FF 	 & 91.3 $\pm 3.5$ & 4.5 & 1.82 $\pm 0.22$ & 1.34E-02 & 1.95 $\pm 0.11$ & 3.46E-03 & 1.060 & 578\\
153SW 	 & 77.7 $\pm 1.8$ & 4.5 & 1.93 $\pm 0.09$ & 2.14E-02 & 1.81 $\pm 0.08$ & 2.46E-03 & 0.971 & 658\\
283FF 	 & 90.1 $\pm 3.5$ & 4.5 & 1.68 $\pm 0.22$ & 0.96E-02 & 1.97 $\pm 0.07$ & 3.90E-03 & 1.116 & 697\\
341FF 	 & 93.4 $\pm 5.4$ & 4.5 & 2.41 $\pm 0.29$ & 1.99E-02 & 2.00 $\pm 0.17$ & 6.60E-03 & 1.109 & 794\\
\noalign{\smallskip}\hline
\end{tabular}
\medskip
\end{table*}

Unfortunately, the  low signal-to-noise high energy BeppoSax data 
 for \pks ~do not further constrain the parameters of the
hot Comptonization component. A fit to the combined MECS and PDS
data with a compTT model resulted in a kT = 47 keV, $\tau = 0.56$ and
a $\chi^2_{\rm red} = 0.97$.
However, the 90\% confidence intervals for these values are 
20.8 keV \lta kT \lta 162 keV and 0.01 \lta $\tau$ \lta 2.85.  
    
The \pks ~fits  at higher energies of either a power law or the hot 
Comptonizing component are
in agreement with the results for the BLS1 fits, as can be seen from the
conversion of the power law slopes to (kT,$\tau$) 
(Titarchuk \& Lyubarskij 1995, eqn 17, 22).
Thus, this component seems to be present in both object classes,
in  NLS1 and BLS1 galaxies.

In addition to this hard component NLS1 galaxies show a second, 
softer Comptonization component. 
This soft component contributes about twice the luminosity  of the
hard component over the 0.1--100 keV energy range and clearly 
dominates the emission in the XMM band. 
As the Compton luminosity of a source is proportional 
L $\sim y \times U \times R^2$ (Haardt, private comm.), 
where U is the radiation energy
density of the soft photons, R the size of the source and 
$y \sim kT\,{\rm max}(\tau,\tau^2)$  is the Compton $y$ parameter,
the emission areas for the two components  with the above
given parameters must be  of similar sizes. 

In contrast to BLS1 galaxies no iron lines from a reflection component could be
detected.  It appears that the `cold reflector' in BLS1 galaxies,
responsible for the iron line emission,  is
replaced by a `warm' (kT $\sim$ 4.5 keV), scattering component.
Further, the higher optical depths imply that any reflection features 
(like the iron line) tend to be more suppressed by Compton scattering in 
the corona itself (Matt et al. 1997).
 
\subsection{Temporal variability: short time scales}
The source shows variability on various time scales. During the 
individual observations we find flux changes of typically $\lta$ 20\%
(the maximum is nearly 50\% in orbit 33) on time scales of 
a few hours.   A closer inspection shows that these flux  variations 
are caused by changes of the soft compTT component only; the hard 
component remains practically constant.
  
\begin{figure}
\psfig{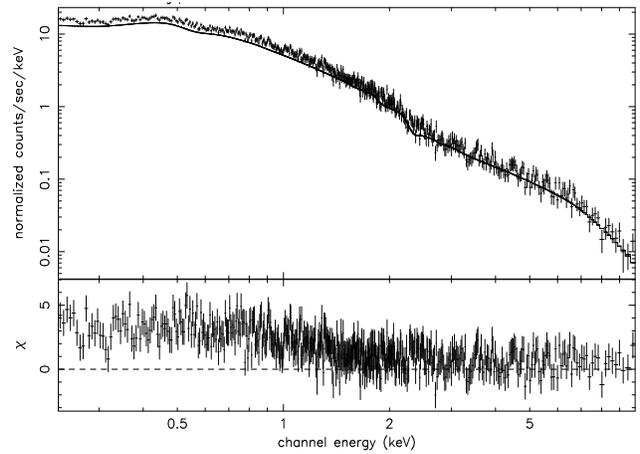}
\caption[]{The high state data of orbit 32 overlaid on a two component
compTT model with parameters  which were obtained from a fit to the
 low state data from the first 5~ksec of the observation.
 The lower panel shows the $\delta \chi^2$ per channel between data and
 this model.}
\label{figure:flare32}
\end{figure}

This is nicely illustrated in 
Fig. \ref{figure:flare32} which shows, as example, the
comparison of the two spectra for the low and high state of orbit 32.
We fitted the first, about 5 ksec,  of the data when the source was at a
low flux level with two compTT models. We then froze the fit parameters and
plot the data from the peak of the light curve and the deviations of
the these data from the model (in $\delta \chi^2$)
in the bottom panel of Fig. \ref{figure:flare32}.
As can be seen, the intensity changes occur only in the soft part of the
spectrum. 
 
\begin{figure}
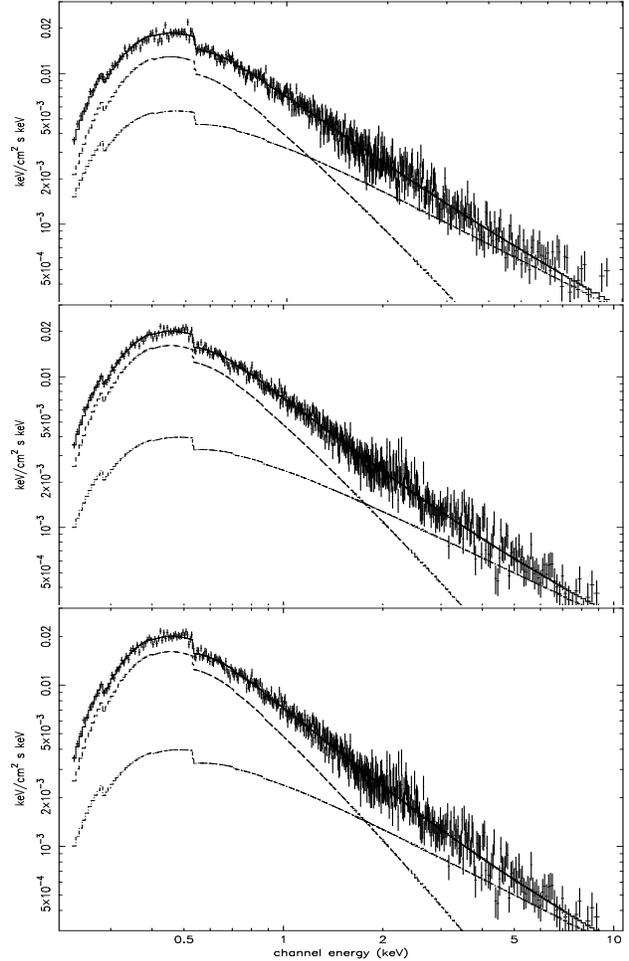

\psfig{figure=H4681F10a.ps,height=4.0truecm,width=8.3truecm,angle=270,%
 bbllx=81pt,bblly=43pt,bburx=531pt,bbury=715pt,clip=}
\psfig{figure=H4681F10b.ps,height=4.0truecm,width=8.3truecm,angle=270,%
 bbllx=81pt,bblly=43pt,bburx=531pt,bbury=715pt,clip=}
\psfig{figure=H4681F10c.ps,height=4.7truecm,width=8.3truecm,angle=270,%
 bbllx=81pt,bblly=43pt,bburx=575pt,bbury=715pt,clip=}
\caption[]{Unfolded spectra from orbit 42 in the preflare(top), 
rise(middle) and flare(bottom) stages. The average count rates for 
the plots, from top to bottom, are 24.2, 25.0, 28.0 counts/s.}
\label{figure:flare42}
\end{figure}

More details of the spectral behavior can be seen in 
Fig. \ref{figure:flare42} which shows the changes 
in the spectrum during orbit 42.
The spectra correspond first to the low count rate state at the
beginning of the orbit, then to the rising phase, and finally to the
high count rate state at the end of this observation.
The spectra were fitted with  a compTT plus a power law model,
 with fixed galactic absorption
 and a fixed temperature of 4.5 keV for the Comptonizing electrons
 to avoid  
 the degeneracy problem between optical depth and electron temperature.
 The model parameters of the power law component remain unchanged within
 the errors while in the Comptonization component the optical depth
 increases slightly with count rate from 2.1 to 2.5. 
 In these fits the increase in the optical depth is responsible 
for the change of the spectral shape: it dominates the spectrum up
 to higher energies as the count rate increases.

The plot further illustrates the difficulties of a hardness ratio analysis:
at low count rates the soft and the hard band are each representing the
contribution of the individual Compton components; at very high
intensities most of the XMM spectral range is dominated by the
soft component.
  
Thus the flux changes occurring on time scales of $\gta$ 2 hours in a relatively
smooth manner are caused by variations of the strength of the soft
compTT component. Whether these variations are  caused by changes of the 
temperature of the Comptonizing gas (at constant scattering depths $\tau$
and normalizations) or by changes of $\tau$ and the normalization  at a 
constant  gas temperature cannot be distinguished from the spectral fits.
The first scenario would require a heating of the same gas volume, 
possibly by electromagnetic processes (i.e. Alfv\'en waves); the latter 
requires changes in the size of the emitting area 
and/or  the volume of the gas responsible for the emission.  
This could be most easily achieved by an increase of the number of 
active emission regions on the disc (Haardt et al. 1994),  which are 
formed at similar temperatures but with initially higher densities  and then
evolve gas dynamically.  
A popular class of models which follows these patterns are
magnetically heated coronas where the energy is primarily stored
in magnetic fields which rise buoyantly from the disc, reconnect 
 and release their energy in flares (Merloni \&  Fabian 2001).
These heat the corona and can trigger an avalanche (Poutanen \& Fabian 1999)
which create in turn a bigger active region.
 
The energy released in a typical burst $\Delta{\rm L} \Delta{\rm t}$ is of 
the order of 10$^{45}$ erg. Using the characteristic size of the 
system (see sects. 4.3, 4.4) of  a few $\times 10^{13}$ cm it turns out that
this energy can be stored easily in a magnetic field of the order 
of $\lta 10^3$ Gauss or, taking the physical parameters from the 
Comptonization fits,  as thermal energy of the gas.
However, recently Merloni \& Fabian (2001) strongly argued in favor 
 of magnetic fields for storing the energy in the corona.  
The time scales for Compton cooling are rather short, of the order of
10$^2$ secs only (Wang et al., 2001), therefore the evolution of the
bursts must be governed by the rate at which energy can be supplied
to the emission region.
 
\subsection{Temporal variability: longer time scales}
We find evidence for a time scale of $\sim$ 24 ksec in the data
from the recurrent occurrence of stronger intensity peaks. 
Unfortunately, the statistical significance for the existence of
such signal cannot be assessed reliably due to the extreme sparsity
of the data.
Although not strictly periodic - there are some orbits where no peak
occurs at the expected time - this time scale seems to represent a quite
stable clock in the system.
It might either be a rather well defined time scale of a dominant 
instability occurring in the disc or it might be related to the
orbital time scale of the emitting matter.

By adopting the empirical broad-line region size versus optical luminosity
relation of Kaspi et al. (2000) and using the H$\beta$ line width of 
1250 km\,s$^{-1}$ (Corbin 1997)  a central black hole mass of 
$4.5\times10^7$ M$_\odot$ was derived for \pks ~by Wang et al. (2001).
Assuming Keplerian motion the above time scale of $\sim$ 24 ks would 
correspond to the orbital period at $\sim$ 3.2 Schwarzschild radii,
i.e. just outside the last stable orbit. 
Considering the uncertainties in the above mass determination this
distance strongly supports the interpretation that the 24 ksec time scale
is related to the orbital motion of the emission region.

\subsection{Temporal variability: the longest time scales}
On even longer time scales, of the order of several satellite orbits, the
source changes by more than a factor of two, but we never see these
transitions to occur during an individual orbit. For example, during
orbits 42 and 45 the source is obviously in a high state which might
thus last for more than a week. 
During the second part of orbit 33, i.e. about 2.5 weeks earlier,
 the flux is more than 2.5 times smaller. 
That these long term intensity variations are a common phenomenon
of NLS1 galaxies is demonstrated in the long light curves of Ark 564 
(Gliozzi et al. 2002) or MCG-6-30-15 (Vaughan et al. 2003).
  
Substantial changes of the radiating area or the physical conditions
of this area, respectively, 
are expected to happen on the dynamical time scale for Keplerian
inflow, $\tau \sim 9~ 10^3~  (\mathrm{r/R}_{\mathrm s})^{3/2}~
({\mathrm M}_{\rm bh}/10^7 ~ {\mathrm M}_\odot) $ sec, where
R$_{\mathrm s}$ is the Schwarzschild radius of the central black hole
 of mass M$_{\rm bh}$.   With the above deduced parameters  for 
\pks ~we find a time scale of $\sim$ 2.3$\times 10^5$s, which is 
consistent with  the observational result that large changes of the
source flux occur on time scales longer than an XMM orbit.

\section{Conclusions}
 
We have analyzed in some detail the observations of \pks ~performed
over nine irregularly spaced XMM orbits spanning a time of more than
two years.
 
The source showed strong variability of $\lta $ 20\% in an individual
observation with typical time scales of  $\gta$ 2 hours. 
The flux variations often exceed the efficiency limit  for
scattering limited spherical accretion. However, as the emission is
 dominated by Compton up-scattered soft photons  in a 
hot corona heated by magnetic reconnection, and the energy is 
stored in the magnetic fields which do not contribute to the 
scattering opacity, the effective efficiency could be as 
large as $\eta \lta B^2/(8\pi\rho c^2) \lta 1$, where $\rho$ is
the mass density of the corona region (Wang et al. 2001).
   
The flaring pattern seems to repeat with a time scale of $\sim$ 24 ksec,
however not in a strictly periodic manner. Interpreting this time 
scale as the orbital time scale of Keplerian motion we find that the
emission occurs at distances of $\gta$ 3.2 Schwarzschild radii from the
central black hole.
On longer time scales,  beyond the temporal spacing of the XMM orbits, 
intensity changes by more than a factor of two are observed, which
can be related to the dynamical time scale of Keplerian inflow
in the disc. 
We do not see any large flare from the source as reported from Ginga
(Remillard et al. 1991) and ASCA (Wang et al. 2001) observations.
Thus that phenomenon must be a rare event.
  
The spectrum of \pks ~(like that of several other NLS1 galaxies as well)
can be well fitted by the sum of two Comptonization models, one
with high temperature (kT $\gta 50$ keV) and low optical depths 
($\tau \sim 0.7$), similar to that found for BLS1 galaxies, and an
additional cooler component (kT $\sim$ 4.5 keV) and larger optical
depths ($\tau \gta 2$).
In contrast to BLS1 galaxies no iron lines from a reflection component could be
detected.  It appears that the `cold reflector' in BLS1 galaxies,
responsible for the iron line emission,  is
replaced by a `warm' (kT $\sim$ 4.5 keV), scattering component.
Flux variations are predominantly caused
by changes of this latter component, whose presence is obviously the 
distinguishing  criterion for the X-ray properties of NLS1 galaxies.

As the spectral capabilities of current instruments are insufficient to
resolve the degeneracy  in the spectral shapes of multiple 
Comptonization model fits, the 
second (cooler) component could equally well be fitted by another
hot component with  extremely small optical depths ($\tau \lta 0.06$).
This would, however, not change the general scenario of a magnetically
very active corona on the accretion discs of NLS1 galaxies but require
different physical parameters for the generation of the coronal 
active regions. 
Unfortunately, current corona models do not seem to have sufficient
predictive power to allow to distinguish between these theoretical 
possibilities.  
For the low mass of the accreting black hole the correspondingly high 
accretion rate is expected to be responsible for higher magnetic
field strengths in NLS1 galaxies (Mineshige et al. 2000) providing a 
mean for storing energy in the corona and leading to strong variability
of the emission from the systems (Merloni \& Fabian 2001, Merloni 2003).
If the turbulent magnetic pressure greatly exceeds that of the gas turbulent
Comptonization might play an important role in producing the soft
X-rays (Socrates et al. 2003). 
This largely unexplored process would provide a direct link 
between details of the disc physics and the observed spectrum. 
  
An unanswered question is related to the radio-loudness of \pks, and
how that affects the emission characteristics. 
The spectral decomposition into two Comptonization components and
the short-term intensity variations are found in other, radio-quiet
NLS1 galaxies as well, therefore we regard this as being a typical NLS1 
characteristic and not related to a possible radio jet, although outflows
are expected from powerful magnetically dominated coronas (Merloni \&
Fabian 2002).
However, the additional emission from a jet might be responsible for
the large flare events, detected by Ginga (Remillard et al. 1991) and 
ASCA (Wang et al. 2001). 
It should be noted that during the flare the spectrum was significantly
harder than in the `quiescence state' (Wang et al. 2001),
 thus indicating an additional, 
 different emission mechanism. 
 
Unfortunately, the X-ray spectra of NLS1 galaxies are nearly featureless and
 thus they bear only little discriminating power to allow parameter
studies for the various 
corona models. The best way to get access to the physical processes 
responsible for the X-ray emission might be, like in BL Lac objects, 
to pursue  detailed temporal analyses of the emission. 

\begin{acknowledgements}
This work is based on observations with XMM-Newton, an ESA science mission
with instruments and contributions directly funded by ESA Member States
 and the USA (NASA). We thank Graziella Branduardi-Raymont for her help
for producing RGS light curves and the referee, P. Uttley, for 
constructive comments which improved the quality of the paper.
 PA \& EF acknowledge support by the International
Max-Planck Research School on Astrophysics (IMPRS).
\end{acknowledgements}

\end{document}